\documentclass[twocolumn,aps,prl,superscriptaddress,showpacs]{revtex4}
\usepackage{hyperref}

\usepackage{graphicx}

\usepackage{amsmath}
\usepackage{epsfig}

\begin{document}
 \title{Universal optical amplification without nonlinearity}

\author{Vincent Josse}
\affiliation{Institut f\"{u}r Optik, Information und Photonik,
Max-Planck Forschungsgruppe, Universit\"{a}t Erlangen-N\"{u}rnberg,
G\"{u}nther-Scharowsky str. 1, 91058, Erlangen, Germany}
\affiliation{LCFIO, UMR8501 du CNRS, B\^{a}timent 503 Centre Universitaire, 91403 Orsay Cedex, France}
\author{Metin Sabuncu}
\affiliation{Institut f\"{u}r Optik, Information und Photonik,  Max-Planck Forschungsgruppe, Universit\"{a}t Erlangen-N\"{u}rnberg, G\"{u}nther-Scharowsky str. 1, 91058, Erlangen, Germany}
\author{Nicolas J. Cerf}
\affiliation{QuIC, Ecole Polytechnique, CP 165, Universit\`e Libre de Bruxelles, 1050 Bruxelles, Belgium}
\author{Gerd Leuchs}
\affiliation{Institut f\"{u}r Optik, Information und Photonik,  Max-Planck Forschungsgruppe, Universit\"{a}t Erlangen-N\"{u}rnberg, G\"{u}nther-Scharowsky str. 1, 91058, Erlangen, Germany}
\author{Ulrik L. Andersen}
\email{andersen@kerr.physik.uni-erlangen.de}
\affiliation{Institut f\"{u}r Optik, Information und Photonik,  Max-Planck Forschungsgruppe, Universit\"{a}t Erlangen-N\"{u}rnberg, G\"{u}nther-Scharowsky str. 1, 91058, Erlangen, Germany}
\date{\today}

\begin{abstract}
We propose and experimentally realize a new scheme for universal
phase-insensitive optical amplification. The presented scheme relies
only on linear optics and homodyne detection, thus circumventing
the need for nonlinear interaction between a pump field and the signal field.
The amplifier demonstrates near optimal quantum noise limited
performance for a wide range of amplification factors.
\end{abstract}


\maketitle

Optical amplification is inevitably affected by
fundamental quantum noise no matter whether it is phase sensitive or phase insensitive as stressed by Louisell \textit{et al.}~\cite{louisell61.pr} and by Haus and Mullen~\cite{haus62.pr}.
The ultimate limits imposed by quantum mechanics on amplifiers was
later concisely formulated by Caves ~\cite{caves82.prd} in
fundamental theorems. This intrinsic noise, intimately linked with
measurement theory and the no-cloning theorem, gives rise to many
inextricable restrictions on the manipulations of quantum states.
For example, microscopic quantum objects cannot be perfectly
transformed, through amplification, into macroscopic objects for
detailed inspection~\cite{glauber}: for phase insensitive amplification nonclassical features of quantum
states, such as squeezing or oscillations in phase space, will be
gradually washed out, and the signal to noise ratio of an information
carrying quantum state will be reduced during the course of
amplification. Despite these limitations, the universal phase
insensitive amplifier is however rich of applications, in particular in
optical communication.
Amplifiers operating at the quantum noise limit are
of particular importance for quantum communication where information
is encoded in fragile quantum states, thus extremely vulnerable to
noise. 

Numerous apparatuses accomplish, in principle, ideal phase
insensitive amplification as, for instance, solid state laser
amplifiers~\cite{shimoda57.jpsj}, parametric
downconverters~\cite{yariv66.qe} and schemes based on four wave
mixing processes~\cite{yuen79.ol}. However, to date phase-insensitive amplification at the quantum limit has been only partially demonstrated~\cite{ou93.prl,levenson93.josab}: 
a number of difficulties are indeed involved in practice, especially
for low gain applications. These difficulties mainly lie in the fact
that the amplified field has to be efficiently coupled, mediated by
a non linearity, to a pump field.

Following a recent trend in quantum information science, where
non~linear media are efficiently replaced by linear optics
\cite{knill01.nat}, we show in this Letter that universal phase
insensitive amplification can also be achieved using only linear
optics and homodyne detection.
The simplicity and the robustness of this original scheme enable us
to achieve near quantum noise limited amplification of coherent
states, even in the low gain regime. 

Let us first briefly summarize the basic formalism describing a
phase-insensitive amplifier \cite{caves82.prd}. Because of the symmetry of such an amplifier,
it can be described by the following input-output transformation:
$\hat{a}_{out}=\sqrt{G}\hat{a}_{in}+\hat{N}$ where
$\hat{a}_{in(out)}$ represent the input (output) annihilation
bosonic operators, $G$ is the power gain and $\hat{N}$ the operator
associated with noise addition. Even for an ideal amplifier, this noise
term must be added to ensure the preservation of the commutation
relations $[\hat{a}_i,\hat{a}_i^\dagger]=1$, and must satisfy
$[\hat{N},\hat{N}^\dagger]=(1-G)$. Thus it can be divided in two
parts: a fundamental quantum part given by
$\hat{N}_q=\sqrt{G-1}\hat{a}_{int}^\dagger$, where $\hat{a}_{int}$
is associated with the unavoidable fluctuations of the internal bosonic mode,
and a scalar classical part denoted $N_{cl}$. Therefore a phase
insensitive amplifier working at the quantum noise limit (i.e. $N_{cl}=0$) obeys the
relation~\cite{caves82.prd}:
\begin{equation}
\hat{a}_{out}=\sqrt{G}\hat{a}_{in}+\sqrt{G-1}\hat{a}_{int}^\dagger
\label{amp_transformation}
\end{equation}
The intrinsic quantum noise, described by $\hat{a}_{int}$, can be
traced back to different physical processes. For instance, spontaneous
emission is unavoidably introduced in laser amplification, whereas,
in parametric amplifiers and four wave mixers, vacuum fluctuations
of the idler mode are added to the output
signal~\cite{shimoda57.jpsj,yariv66.qe,yuen79.ol}. In Raman
amplifiers and Brillouin amplifiers zero-point fluctuations of
respectively lattice vibrational modes (optical phonon) and acoustic
phonon modes cause the noise~\cite{louisell61.pr}.

The efficiency of a phase-insensitive amplifier is typically quantified by the noise
figure~\cite{caves82.prd,levenson93.josab}, which is defined by
NF$\equiv$ SNR$_{out}$/SNR$_{in}$. Here SNR$_{in(out)}$ is the
signal-to-noise ratio of the input (output) field. For coherent state amplification, the noise figure then reads
\begin{equation}
NF=\frac{G}{2G-1+\Delta^2N_{cl}}
\label{NF}
\end{equation}
which is maximized for quantum noise limit operation corresponding
to $\Delta^2N_{cl}=0$. However, provided that the spurious technical
noise $N_{cl}$ is constant or has only a weak dependence on $G$, the noise figure still approaches the
quantum limit of 3dB in the high gain regime. The situation is
different at low gains, as technical noise or internal losses become
devastating for quantum noise limited
performance~\cite{ou93.prl,levenson93.josab}. To date, these effects
have hitherto prevented the full demonstration of quantum noise
limited phase-insensitive amplification in the low gain
regime~\cite{note2}, which is the domain of interest in the context
of quantum information science.

For sake of completeness we mention that another class of amplifiers characterized by phase sensitive operation allows for noiseless amplification, provided the analysis is restricted to just one quadrature~\cite{caves82.prd,levenson93.josab}. Such an amplifier is described by the relation: $\hat{a}_{out}=(1+G)/\sqrt{4G}\hat{a}_{in}+(1-G)/\sqrt{4G}\hat{a}_{in}^\dagger$ and the noise figure is: NF=1.  

\begin{figure}[h] \centering \includegraphics[width=7.5cm]{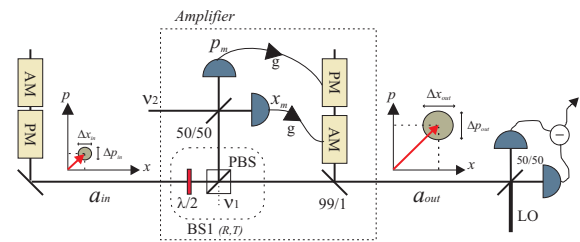} \caption{\it Conceptual diagram
 of the amplifier setup. The gain is determined by the transmission of the equivalent beam splitter (BS), (realized here by a combination of a half wave plate and a polarizing beam
 splitter (PBS)). AM and PM are amplitude an phase
 modulators respectively. Also shown are the phase space diagrams of the input coherent state and the amplified output state.} \label{setup}  \end{figure}

We now show that the amplifier transformation
(Eq.~\ref{amp_transformation}) can be realized using only linear
optics, homodyne detection and feedforward, rendering the complex
coupling between a strong pump and the signal inside a nonlinear
crystal superfluous. Our scheme is illustrated schematically inside
the dashed box in Fig.~\ref{setup}, and runs as follows. The input
signal, represented by $\hat{a}_{in}$, is impinging on a beam
splitter with transmission $T$ and reflectivity $R$, and hence
transformed into
$\hat{a}_{in}'=\sqrt{T}\hat{a}_{in}-\sqrt{R}\hat{v}_1$ where the
annihilation operator $\hat{v}_1$ represents the vacuum mode
entering the dark port of the beam splitter. Conjugate quadrature
amplitudes, e.g. the amplitude $\hat{x}=\hat{a}+\hat{a}^\dagger$ and the phase quadrature
$\hat{p}=-i(\hat{a}-\hat{a}^\dagger)$, are simultaneously measured on the reflected part by
dividing it on a 50/50 beam splitter and subsequently performing
homodyne measurements on the two output beams. The measured
quadratures are
\begin{eqnarray}
\hat x_m=\frac{1}{\sqrt{2}}(\sqrt{R}\hat x_{in}+\sqrt{T}\hat x_{v1}+\hat x_{v2})\\
\hat p_m=\frac{1}{\sqrt{2}}(\sqrt{R}\hat p_{in}+\sqrt{T}\hat p_{v1}-\hat p_{v2})
\end{eqnarray}
Here $\hat{x}_{v(1,2)}$ and $\hat{p}_{v(1,2)}$ denote
the quadratures of the uncorrelated vacuum modes entering at the two beam
splitters. These projective measurements (with outcomes represented
by their eigenstates and corresponding eigenvalues $x_m$ and $p_m$)
are then used to control a unitary displacement operation on the
remaining system~\cite{bjork88.pra}. The feedforward loop can be described without any measurement by
considering the unitary operator,
$\hat{D}=\exp(g\hat{x}_m(\hat{a}_s-\hat{a}_s^\dagger))\exp(ig\hat{p}_m(\hat{a}_s+\hat{a}_s^\dagger))$,
where $g$ is the electronic gain and then subsequently
tracing out the "measured" system. This
results in the following transformation: $\hat{a}_s'\rightarrow
\hat{D}^\dagger \hat{a}_s'
\hat{D}=\hat{a}_s'+g(\hat{x}_m+i\hat{p}_m)/2$ and by choosing an
electronic gain of $g=\sqrt{2R/T}$ we arrive at
\begin{eqnarray}
\hat{a}_{out}=\sqrt{\frac{1}{T}}\hat{a}_{in}+\sqrt{\frac{1}{T}-1}\hat{v}_2^\dagger
\label{transformation}
\end{eqnarray}
Setting $G=\frac{1}{T}$, we exactly recover the transformation for
an ideal phase-insensitive amplifier given by
(\ref{amp_transformation}), where the amplification factor is
controlled by the beam splitting ratio. Note that the noise that enters from the vacuum fluctuations on $\nu_1$
is automatically cancelled out in the output via the feedforward. We also note that a related scheme, where noise entering a beam splitter was cancelled via feed forward, was used in ref.~\cite{lam98.prl} to build a noiseless amplifier (with NF=1 for the amplitude quadrature). However, in contrast to our proposal, this scheme, apart from being phase sensitive, was not fully operating at the fundamental quantum limit~\cite{caves82.prd}. A truly quantum noise limited phase sensitive amplifier 
based on the same principles was recently proposed~\cite{filip05.pra}, but it requires 
a nonclassical resource, namely a squeezed vacuum state. 
 
Interestingly, the fundamental amplifier noise, represented by $\hat{\nu}_2$, arises
from the vacuum fluctuations that enter through the dark port of the
50/50 beam splitter used for $\hat{x}$ and $\hat{p}$ quadrature
measurements. The amplifier noise is therefore directly related to the noise penalty
associated with simultaneous measurement of conjugate quadratures.
The close link between amplification and measurement theory
\cite{caves82.prd} is thus particularly emphasized by our scheme.

The amplifier proposed in this Letter is phase-insensitive, and, in principle, amplifies any input state at the
quantum limit. In the following, we demonstrate experimentally
the amplification of a particular quantum state, namely the
coherent state. The experimental setup is shown in Fig.~\ref{setup}.
The laser source was a monolithic continuous wave Nd:YAG laser at
1064~nm. A small part of the laser beam was tapped off to serve as
an input signal to the amplifier and the rest was used as local
oscillator beams. Since the output from a laser is not a perfect
coherent state due to low frequency technical noise, we define our
coherent state to reside at a certain sideband frequency which we
chose to be 14.3~MHz, within a bandwidth of 100kHz. At this
frequency the laser was found to be shot noise limited, and by
applying modulations at 14.3~MHz (by independently controlling an
amplitude and a phase modulator), the sidebands are excited and thus
serve as a perfect coherent state.

The coherent state is then directed to the amplifier where it is
divided by a beam splitter; the reflected part is measured and the
transmitted part is displaced according to the measurement outcomes.
Simultaneous measurements of the amplitude and phase quadrature are
performed by combining the reflected signal beam with an auxiliary
beam, $\nu_2$, with a $\pi /2$ phase shift and balanced intensities. The sum
and difference of the photocurrents generated by two high quantum
efficiency photodiodes then provide the simultaneous measurement of
amplitude and phase quadrature (this strategy is not shown
explicitly in the figure). The outcomes are sent to electronic
amplifiers with appropriate gains and then finally fed into
independent modulators. The modulators are placed in an auxiliary
beam which is coupled to the signal beam via an asymmetric beam
splitter which transmits 99\% of the signal and reflects 1\% of the auxiliary beam, thus leading to a negligible small noise addition. After displacement, the amplified signal is directed
into a homodyne measurement system for verification.

The performance of the system is characterized by measuring the
spectral noise properties of the signal before and after
amplification. Since the quadrature statistics of the involved
fields are Gaussian, measurements of the first
($\langle\hat{x}(\hat{p})_{out}\rangle$) and second moments
($\Delta^2\hat{x}(\hat{p})_{out}$) of two conjugate observables,
such as the amplitude and phase quadrature, suffice to fully
characterize the states. Both quadratures are measured at the
sideband frequency using a standard homodyne detection techniques.
To ensure consistent comparison between the input and output signal,
these measurements are realized by the same homodyne detector.

\begin{figure}[h] \centering \includegraphics[width=7.5cm]{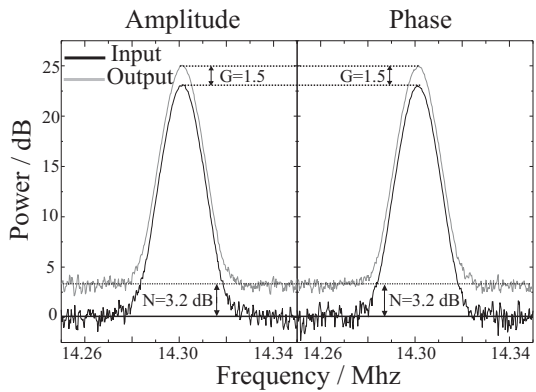} \caption{\it Power spectra showing the operation
of the amplifier for conjugate quadratures. The beam splitter was
set to 1:2 enabling an optical gain of $G=1.5 (1.8dB)$. The mean
value of the field is amplified by 1.8dB and the noise is
consequently increased by 3.2dB, which is very close to the ideal
noise level of 3dB above the shot noise. The noise figure is $NF=0.7$. The resolution
bandwidth is 10kHz and the video bandwidth is 30Hz.}
\label{amplifier}  \end{figure}

An example of a specific amplifier run is shown in Fig.
\ref{amplifier}. Here we set the beam splitting ratio to 1:2 in
order to reach an optical gain of 1.5. The spectral densities of the
amplitude and phase are shown over a 100~kHz frequency span for the
input signal and the amplified output signal. Considering the whole
span as a part of the quantum state, the heights of the peaks
correspond to the coherent mean values whereas the noise floor can
be regarded as the actual noise in the state. Therefore the
amplification factor, which is roughly the difference between the
input and output peaks, as well as the added noise, which is the
difference between the shot noise limit and the noise floor, can be
easily estimated. It is evident from the plots that additional noise
has been added to the signal as a result of the amplification
process.

\begin{figure}[h] \centering \includegraphics[width=7.5cm]{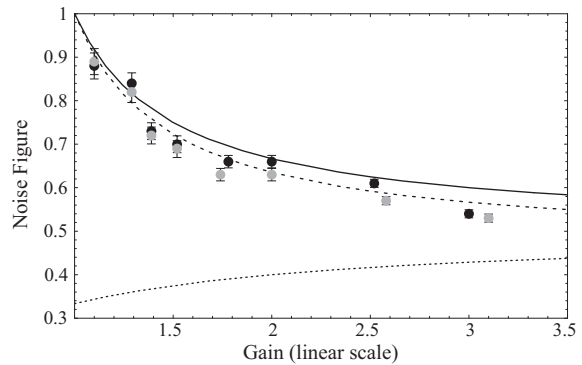} \caption{\it The noise figure, NF, as
a function of the gain, G. The black (grey) dots represent the
experimental data for the amplitude (phase) quadrature. The solid
line represent the quantum noise limit (given by Eq.~\ref{NF}),
whereas the predicted noise figure for our device with imperfect
detectors is shown by the dashed line. For comparison, the dotted
line corresponds to an amplifier with two vacuum units of extra
technical noise. Errors mainly stem from the inaccuracy in
determining the quantum efficiency of the photodiodes.}
\label{noisefigure}
\end{figure}

To evaluate the noise figure, we estimated accurately the
gain and the added noise at 14.3~MHZ. This was realized in a zero
span measurement over 2 seconds by subsequently switching on and off
the modulation. Moreover, to avoid erroneously underestimation of
the noise power, all the measurement have been corrected for losses
occurring in the homodyne detection. The total efficiency, including
mode matching and photodiode quantum efficiency, has been carefully
estimated to $\eta_{hd}=0.83$.

In Fig.\ref{noisefigure}, we report the noise figure of our
amplifier for a whole range of gains (corresponding to different
transmission coefficients and optimized electronic gains). By comparing the
experimental results with the ideal ones (calculated from Eq.(\ref{NF}) and indicated by the solid line),
we clearly see that the amplifier operates close to the fundamental
limit even for low amplification factors. The small deviation to the
ideal amplifier performance is due to imperfections in the in-line
homodyne detector and feedforward electronics. These limiting factors were partly overcome by paying special attention to the construction and alignment of the system. The efficiency of the homodyne detector amounted to 93\% (95\% photodiode efficiency and 99\% mode overlap efficiency) and the electronic noise of the detectors was overcome by using newly designed ultra sensitive detectors. Taking these imperfections into account the theoretically expected noise figure is given by $NF=\eta G/(2G-2+\eta)$ where $\eta$ is the overall efficiency of the detector system.
This expression, which tends to the limit
$NF_{l}=0.46 \;(-3.3dB)$ for high gains, is shown in Fig.
\ref{noisefigure} by the dashed line: it is in good agreement with the experimental results, demonstrating that basically no additional
technical noise is invading the amplifying process.

The challenge of realizing such a quantum noise limited amplifier in
the low gain regime is highlighted by considering the behavior
of an amplifier that exhibits only two vacuum units of extra
technical noise ($\Delta^2 N_{cl}=2$ in Eq.~(\ref{NF})). As
mentioned earlier and clearly illustrated by in
Fig.~\ref{noisefigure}, even such a small amount of background noise,
which is quite common for amplifiers, leads to a strong deviation
from the quantum noise limit at low gains.

\begin{figure}[h] \centering \includegraphics[width=7cm]{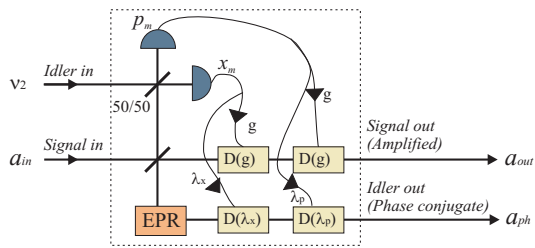}
\caption{\it Proposed scheme for a phase conjugating amplifier with
the nonlinearity put off-line. The displacements, indicated by D, can be performed as shown in Fig. 1.} \label{phaseconjugate}
\end{figure}

To complete the investigation of the system, we finally focus here
on the existence of the phase conjugate amplified output state. This
state, mirrored about the amplitude quadrature axis in phase space
with respect to the input state, must be present in all amplifier to
ensure unitarity~\cite{caves82.prd}. In downconverters, this mode
is the idler output and thus easily accessible for further
processing. However it is not always directly accessible: e.g. in a laser
amplifier this mode is scattered into vibrational modes of the
atoms. But where
is the phase conjugate output in our scheme? It turns out that it can be
extracted by the introduction of an entangled ancilla, as shown in
Fig.~\ref{phaseconjugate}. The amplifier settings are not changed
(so equation (\ref{amp_transformation}) still holds), but now, in
addition, one half of the entangled ancilla is injected into the
empty port of the variable beam splitter and the other half is
displaced according to the classical measurement outcomes. The amplification noise is not affected by this
since the noise due of the entangled ancilla is cancelled out, as mentioned
earlier. The
electronic gains of the classical currents before displacement are
$\lambda _x=\sqrt{2/T}$ and $\lambda _p=-\sqrt{2/T}$ for the
amplitude and phase quadrature, respectively. For perfect
entanglement in the ancilla we find the following input-output
relation for the additional output mode:
\begin{equation}
\hat{a}_{ph}=\sqrt{\frac{R}{T}}\hat{a}_{in}^\dagger+\sqrt{\frac{1}{T}}\hat{v}_2
\label{pc}
\end{equation}
Eq. (\ref{transformation}) and Eq. (\ref{pc}) mimic the ones of a
downconverter and allow us to interpret $\hat{a}_{in}$ and
$\hat{a}_{out}$ as being the input and output signal modes and
$\hat{v}_2$ and $\hat{a}_{ph}$ as being the input and output idler
modes.

In conclusion, we have proposed and experimentally demonstrated that
a phase-insensitive amplifier can be constructed from simple linear
optical components, homodyne detectors and feedforward. Quantum
noise limited performance was exhibited, in particular at low
gains, only limited by inefficiencies of the in-line detection
process. The fact that our amplifier exhibits nearly quantum noise
limited performance at low gains suggests that it can be used to
amplify non classical states (such as squeezed states and Schrodinger
cat states) and still maintain some of their nonclassical features
such as squeezing and interference in phase space. Furthermore, we
believe that such an amplifier can find usage in the field of
quantum communication, where optimal amplification of information
carrying quantum states is needed partly to compensate for down
stream losses of a quantum channel and partly to enable an arbitrary
quantum cloning function~\cite{braunstein}.
One particular cloning transformation of a coherent state was recently demonstrated with a fixed gain amplifier~\cite{andersen05.prl}. 

We thank Norbert L\"{u}tkenhaus and Radim Filip for fruitful
discussions. This work has been supported by the EU project no.
FP6-511004 COVAQIAL.

\end{document}